\documentclass[12pt,a4paper]{article}

\begin{document}

\def\be{\begin{equation}}
\def\ee{\end{equation}}
\def\tr{{\rm tr}}
\def\nn{\nonumber}
\def\bk{\bar{\kappa}}
\def\la{{\lambda}}
\def\al{{\alpha}}
\def\we{\wedge}
\def\ga{{\gamma}}
\def\de{\delta}
\def\om{\omega}
\def\Om{\Omega}
\def\pr{{\pi\rho}}
\def\xj{x^{11}}
\def\f#1{f^{(#1)}}
\def\h#1{h^{(#1)}}
\def\Z{{Z\!\!\!Z}}
\def\oM{{\overline M}}
\def\sgn{{\rm sgn}}
\def\sm{$ {\rm T}_{\rm SM} $}
\def\bea{\begin{eqnarray}}
\def\eea{\end{eqnarray}}
\def\ba{\begin{array}}
\def\ea{\end{array}}
\def\pa{{\partial}}

%%%%%%%%%%%%%%%%%%%%%%%%%%%%%%%%%%%%%%%%%%%%%%%%%%%%%%%%%%%%%%%%%

\begin{titlepage}
\begin{flushright}   IFT-7/2000  \end{flushright}
\begin{flushright}   hep-th/0003233  \end{flushright}

\vspace{1cm}

\centerline{\LARGE{\bf {Anomaly cancellation in M--theory}}}
\vskip .5cm
\centerline{{\LARGE{\bf on orbifolds}}}

\vskip .5cm

\vspace{1cm}

\centerline{\bf Krzysztof A. Meissner and Marek Olechowski}

\vskip .5cm
\centerline{\em Institute of Theoretical Physics}
\centerline{\em Warsaw University}
\centerline{\em Ho\.za 69, 00-681 Warszawa, Poland}

\thispagestyle{empty}

\vspace{2cm}

\begin{abstract}

We present calculation of the anomaly cancellation in M--theory on 
orbifolds $S^1/\Z_2$ and $T^5/\Z_2$ in the upstairs approach.  
The main requirement that allows one to uniquely define solutions to
the modified Bianchi identities in this case is that the field
strength $G$ be globally defined on $S^1$ or $T^5$ and properly
transforming under $\Z_2$. We solve for general $G$ that satisfies
these requirements and explicitly construct anomaly--free theories in
the upstairs approach. We also obtain the solutions in the presence of
five--branes. All these constructions show equivalence of the
downstairs and upstairs approaches. For example in the $S^1/\Z_2$ case
the ten--dimensional gauge coupling and the anomaly cancellation at
each wall are the same as in the downstairs approach.   

\end{abstract}

\vfill
\noindent
March 2000

\end{titlepage}

%%%%%%%%%%%%%%%%%%%%%%%%%%%%%%%%%%%%%%%%%%%%%%%%%%%%%%%%%%%%%%%%%%%%%

%
%
\section{Introduction}

Dualities in string theory connect seemingly different theories. One
of the most unexpected predictions in this web of dualities 
is the appearance of 11--dimensional theory (M--theory) that manifests
itself at low energies as 11--dimensional supergravity. The theory
compactified on different spaces is dual to different string theories.
Compactification on $S^1/\Z_2$ gives the M--theoretic extension of the
heterotic $E_8\times E_8$ string, i.e. a geometric picture of
11--dimensional spacetime with two 10--dimensional walls at
the ends of a finite interval along eleventh dimension
\cite{HW1,HW2,Witten}. While the supergravity multiplet (graviton,
gravitino and antisymmetric tensor fields) can penetrate in the $d=11$
bulk, the two $E_8$ gauge supermultiplets are confined to the two
walls, respectively.  
The requirement of anomaly cancellation gives the relation 
between the 11--dimensional Newton's constant $\kappa$ and the
10--dimensional gauge coupling constant $\lambda$ -- it was 
first obtained in \cite{HW2} with a numerical factor of 2 missing. The
correct relation was first obtained in \cite{Conrad} and subsequently
by many authors both in the downstairs and in the upstairs approaches
\cite{repeat,flo}.
However, as was noted in \cite{bds}, the previous calculations in the 
upstairs approach were inconsistent because of incorrect taking into
account the $\Z_2$ symmetry. Rather surprisingly, the result of the
upstairs calculation of \cite{bds} gave different result than the
downstairs calculation of \cite{Conrad}. 

Another interesting example of an orbifold in M--theory
compactification is $T^5/\Z_2$ \cite{witfb,dasm} which is 
dual to string theory IIB compactified on $K3$. There
are 32 fixed six--planes and it turns out that there must be 16 
additional twisted sector multiplets living on these six--planes to
cancel anomalies. In the presence of five--branes some of these tensor
multiplets are transferred to the branes. 

In section 2 we analyze the anomaly cancellation in the
upstairs approach for the compactification on $S^1/\Z_2$ 
and the results show complete equivalence of the downstairs and the
upstairs approaches.  
The gauge anomaly is restricted to the walls and
there are no mixing terms between the walls so that 
the cancellation proceeds exactly as in the downstairs approach. 
The combination $\lambda^6/\kappa^4$ turns out to be exactly the same
as in \cite{Conrad} (while in ref.\ \cite{bds} it differs by a factor
of 3). 
In section 3 we include five--branes and derive the
field configuration for which the total anomaly vanishes. 
In section 4 
we explicitly construct a theory compactified on $T^5/\Z_2$
and show that there exists in the upstairs approach a configuration of
fields that is locally (at each fixed six--plane) anomaly--free. 
Section 5 presents some conclusions.

\section{M--theory on $S^1/\Z_2$}

The low energy limit of M--theory compactified on $S^1/\Z_2$ was
given  
in \cite{HW1,HW2}. We will use in this paper slightly different
normalization of the fields (the same as in \cite{bds}).  
The actions for the theory in the upstairs and downstairs
approaches look the same but the range of integration and the
normalizations are different.

In the upstairs approach we have full circle (of length $2\pi\rho$)
in the eleventh dimension with two branes located at $x^{11}=0$ and
$x^{11}=\pi\rho$ 
and we impose $\Z_2$ symmetry on fields (for example $C_{ABC}$ and
$G_{ABCD}$ with $A,B,C,D=1\ldots10$ are odd under $\Z_2$ and therefore
vanish on the walls while $C_{11AB}$ and $G_{11ABC}$ are even).
The action reads (for simplicity we keep only bosonic terms)
\be
S=-\frac{1}{2\bk^2}\int_{\oM_{11}}\sqrt{-g}\left[
R+\frac{1}{48}G^2\right]
-\frac{1}{12\bk^2}\int_{\oM_{11}}
C\we G\we G
-\frac{1}{4\la^2}\sum_i \int_{M_{10}^{i}}\sqrt{-g}F^2_i\,.
\label{act}
\ee
In the downstairs approach we take an interval (of length $\pi\rho$)
with the two walls on its ends and instead of the $\Z_2$ symmetry we
have to impose appropriate boundary conditions.
The action is the same but we have to
replace $\oM_{11}$ by $M_{11}$ (full circle by the interval in the
eleventh dimension) and the coupling constant $\bk$ by $\kappa$ where:
\be
\kappa^2=\frac12\bk^2,\ \ \ \ \
\int_{M_{11}}=\frac12\int_{\oM_{11}}\,.
\label{updown}
\ee

Anomaly cancellation requires modifications of the Bianchi
identities that involve sources on the walls. We will solve below for
general $G$ and $C$ that satisfy these identities and are 
well defined on the full circle (i.e. are periodic). 

Let us start with some definitions that will prove useful later on.
We introduce a space $\Phi$ of differentiable functions defined on
the interval $[0,\pr]$ and satisfying
\be
g\in \Phi: \ \ \ \ g(0)=1,\ \ \ g(\pr)=0\,.
\label{gcond}
\ee
For any $g\in \Phi$ we define two periodic functions, $f_g^{(1)}$ and
$f_g^{(2)}$, defined on the circle $(-\pr,\pr]$:
\bea
\f{1}_g(\xj)\!\!&=&\!\!\sgn(\xj)g(|\xj|)\,,\nn\\
\f{2}_g(\xj)\!\!&=&\!\!\sgn(\xj)(g(|\xj|)-1)\,,\\
\f{1}_g(\pr)\!\!&=&\!\!\f{2}_g(\pr)=0\,.\nn
\eea
These functions  constitute the most general $\Z_2$--odd primitives of
delta functions located, respectively, at 0 and $\pr$, having no other
singularities and defined globally on the circle. 
The derivatives of these functions are:
\be
\frac{\pa\f{i}_g}{\pa\xj}=2\de^{(i)}+h_g
\label{fder}
\ee
where $h_g(\xj)=g'(|\xj|)$ is regular everywhere and $\de^{(i)}$ is
the Dirac delta function located at the $i$--th wall. Later we will
use the same symbol to denote the corresponding one form. It is easy
to
prove that for any $g$ 
\be
\int_{S_1}d\xj h_g(\xj)\f{i}_g(\xj)\f{j}_g(\xj)=\frac13-\de_{ij}\,.
\ee
Regularization of the delta function gives also:
\bea
\de^{(i)} \f{j}_g\f{k}_g&\to &\frac13(\de_{ij}\de_{ik})\de^{(i)}\,,
\label{dff}\\
\de^{(i)}\f{j}_g&\to &0
\label{usres}
\eea
(in sense of distributions i.e. when integrated with regular
functions).

Having defined the primitives of delta functions we can now solve for
$G$ in the modified Bianchi identities. They read 
\be
dG=-\ga\sum_i \de^{(i)}\we I_i
\label{bianid}
\ee
where $\ga=(4\pi)^2\bk^2/\la^2$ and
\be
I_i=\frac1{4\pi^2}\left.\left(\tr F_i^2-\frac12 \tr
R^2\right)\right|_{M_{10}^i} \,.
\ee

An integral of $dG$ over a $C_4\times I$ where the interval $I$
surrounds only one wall is not vanishing but the integral over the
entire interval should vanish for $G$ to be well defined. Therefore
the sum of $I_1$ and $I_2$ must be cohomologically trivial:
\be
\int_{C_4}(I_1+I_2)=0\,.
\label{iglob}
\ee

Because of (\ref{iglob}) there exists a form of $G$ that is explicitly
well defined globally. We can write
\bea
I_1&=&H_4+d\Om_1\,,\nn\\
I_2&=&-H_4+d\Om_2
\label{ih4}
\eea
where $H_4$ is a harmonic 4-form and $\Om_i$ are 3--forms well defined
globally. Then we can write the solution of (\ref{bianid}) in the form
\be
G=d\tilde{C}-\frac{\ga}{2}\sgn(\xj)H_4+\ga\sum_i \de^{(i)}\we \Om_i
\label{gglob}
\ee
where each of the terms is well defined globally. Unfortunately, 
we do not know the explicit form of (\ref{ih4}) and 
it is difficult to connect $\tilde{C}$ with the $C$ field from
M--theory where $G=dC$ in the bulk. Therefore we will seek the
solution of (\ref{bianid}) in a different form which is only
implicitly well defined globally. The solution to (\ref{bianid})
satisfying $G=dC$ in the bulk is given by 
\be
G=dC+\ga\sum_i \de^{(i)}\we \om_i
\label{gsol}
\ee
where $I_i=d\om_i$ locally\footnote{In general, $\om_i$ is not well
defined globally since $I_i$ can have contributions from harmonic
forms ($H_4$ in (\ref{ih4})) but $C$ is likewise not well defined
globally and the two contributions should cancel to produce well
defined $G$ like in (\ref{gglob}).}. 
In order to analyze the consequences of $\Z_2$ symmetry, 
we will take care to define all forms globally at least in the
eleventh dimension. We expect $G$ to be a regular form and therefore
we expect that singularities in $\de C$ should cancel $\de^{(i)}$ in
(\ref{gsol}). Let us define a regular three--form $C_{reg}$ (depending 
on functions $g_1$ and $g_2$) by the relation
\be
C=C_{reg}-\frac{\ga}{2}\sum_i \f{i}_{g_i}\om_i\,.
\label{creg}
\ee
Substituting this form to (\ref{gsol}) we get a regular (along $\xj$)
form 
\be
G=dC_{reg}-\frac{\ga}{2}\sum_i h_{g_i}d\xj\we\om_i
-\frac{\ga}{2}\sum_i \f{i}_{g_i} I_i
\label{gdc}
\ee
where we used (\ref{fder}).

Integration of Bianchi identities over $I\times C_4$ where $I$ is an
interval along $\xj$ comprising only one wall shows that $G$ in
(\ref{gdc}) is globally well defined only if 
\be
g_1(\xj)=g_2(\xj)=g(\xj)\,.
\ee
Hence, the same function has to be used in $\f{1}_g$ and $\f{2}_g$
(and consequently there is only one regular derivative
$h$). Therefore, we will suppress subscript $g$ from now on.

To calculate all contributions to the anomaly, we have first to find
gauge variation of $C$. Starting from the condition $\de G=0$ and
using equations (\ref{gsol}) and (\ref{creg}) we get
\be
\de C_{reg}=dB_{reg}-\frac{\ga}{2}h \sum_i d\xj\we\om_i^1
\label{decreg}
\ee
where $\de\om_i=d\om_i^1$.

Let us now calculate the anomaly in the upstairs approach.
The contribution from the topological term in the action is equal to
\be
\de S_{top}=-\frac{1}{12\bk^2}\int\de C\we G\we G\,.
\label{destop}
\ee
The relation $G=dC$ valid in the bulk requires $C$ (and not $C_{reg}$) 
in the above expression -- we would break supersymmetry in the bulk
otherwise. Let us note that the authors of \cite{bds} used $\de
C_{reg}$ in (\ref{destop}) and not $\de C$ and it is one of the
reasons for the difference between the results of the present paper
and that of \cite{bds}\footnote{In \cite{bds} a linear function was
chosen for $g$ but as we show in the present paper nothing really
depends on the choice of $g$ as long as conditions (\ref{gcond}) are
satisfied.}.     

Using the expressions (\ref{gdc}) and (\ref{decreg}) we have
\bea
\de C&=&dB_{reg}-\frac{\ga}{2} h \sum_i d\xj\we\om_i^1
-\frac{\ga}{2}\sum_i \f{i}d\om_i^1\,,
\label{dec}\\
G&=&dC_{reg}-\frac{\ga}{2}  h \sum_i d\xj\we\om_i
-\frac{\ga}{2}\sum_i \f{i} I_i\,.
\label{deg}
\eea
It is easy to show (see relation (\ref{usres})) that the terms with
$dB_{reg}$ and $dC_{reg}$ do 
not contribute to (\ref{destop}) due to regularity of $B_{reg}$ and
$C_{reg}$ and the fact that $G_{ABCD}|_i=0$. Therefore we get
\bea
\de S_{top}&=&\frac{1}{12\bk^2}\left(\frac{\ga}{2}\right)^3
\int_{\oM_{11}}
\sum_{ijk}\left(h d\xj\we\om_i^1+\f{i}d\om_i^1\right)\we\nn\\
&&\ \ \ \ \ \ \ \ \ \ \ \ \ \ \ \ \ \ \ \ \ \ \ \ \ \ 
\left(h d\xj\we\om_j+\f{j}I_j\right)\we
\left(h d\xj\we\om_k+\f{k}I_k\right)\nn\\
&=&\frac{\ga^3}{96\bk^2}
\sum_{ijk}\left(\int_{S_1}h\f{j}\f{k}\int_{M_{10}}\om_i^1\we I_j\we
I_k \right.\\
&&\left. \ \ \ \ \ \ \ \ \ \ \ \ \ 
-2\int_{S_1}h\f{i}\f{k}\int_{M_{10}}d\om_i^1 \we \om_j\we I_k
\right)\nn\\
&=&\frac{\ga^3}{96\bk^2}
\sum_{ijk}\int_{M_{10}}\left[\left(\frac13-\de_{jk}\right)\om_i^1 \we
I_j \we I_k
+2\left(\frac13-\de_{ik}\right)\om_i^1 \we I_j \we I_k\right]\,.\nn
\eea
Expanding this result shows that the terms which mix contributions
from different walls cancel and we have the final result
\be
\de S_{top}=-\frac{\ga^3}{48\bk^2}
\sum_{i}\int_{M_{10}}\om_i^1\we I_i\we I_i\,.
\label{destopf}
\ee

The next contribution to the anomaly is the Green--Schwarz term -- we 
take it to be of the form\footnote{Choosing instead $G\we X_7$ with
$X_8=dX_7$ would give the same contribution to the anomaly expressed
as 12--form but the anomaly cancellation in the 10--form would then 
require a local counterterm.} \cite{HW2,gsterm}
\be
S_{GS}=-\frac{1}{\ga}\int_{\oM_{11}} C\we X_8\,.
\label{sgs}
\ee
Calculating the gauge variation of this term using (\ref{dec}) we get 
\bea
\de S_{GS}&=&\frac12\int_{\oM_{11}}\sum_i(h d\xj\we\om_i^1+
\f{i}d\om_i^1)\we X_8\nn\\
&=&-\int_{\oM_{11}}\sum_i \de^{(i)}\we\om_i^1\we X_8
=-\sum_i\int_{M_{10}}\om_i^1\we X_{8,i}\,.
\label{desgs}
\eea

The third contribution to the anomaly is the one--loop result -- it is 
given by  
\be
\de S_{1-loop}=\frac{\pi}{3}
\sum_{i}\int_{M_{10}}\om_i^1\we I_i\we I_i+
\sum_{i}\int_{M_{10}} \om_i^1\we X_{8,i}\,.
\label{desonel}
\ee

The requirement of vanishing of the total anomaly i.e the sum of
(\ref{destopf}), (\ref{desgs}) and (\ref{desonel}) gives
\be
\frac{\la^6}{\bk^4}=\frac{(4\pi)^5}{4}\,.
\ee
Translating this result to the downstairs language using
(\ref{updown}) we get 
\be
\frac{\la^6}{\kappa^4}=(4\pi)^5
\ee
and this relation is exactly the same as in \cite{Conrad}.

\section{M--theory on $S^1/\Z_2$ with five--branes}

Let us now include five--branes -- 
possible non--perturbative objects in M--theory that couple
magnetically to $G$ and act as sources for modified Bianchi
identities. Their presence can in general modify the condition
(\ref{iglob}) and therefore require non--standard identification of
gauge fields with the connection. We would like to show here that we
can repeat the whole discussion in the presence of five--branes in the
upstairs approach. 

Let us take into account the five--branes parallel to three space
dimensions and two CY dimensions (it is possible that with the
non-standard embedding it is no longer a Calabi--Yau space but let us
leave this subtlety aside). These five--branes act then as additional
sources in the Bianchi identity 
\be
dG=-\ga\sum_i \de^{(i)}\we I_i-\ga\sum_{\al}\de^{(\al)}_5
\label{brbianid}
\ee
where $\de^{(\al)}_5$ are products of five delta one forms along four
CY dimensions and $\xj$. The requirement that $G$ be globally defined 
gives  
\be
\int_{C_4}\left(I_1+I_2\right)+[C_\al]=0
\label{BIglob}
\ee
where $[C_\al]$ is equal to the number of five--branes surrounded by a
given cycle $C_4$.

In analogy to the previous case let us introduce a $\Z_2$ odd and
periodic function with jumps at $\xj=x_{\al^+}>0$ and $\xj=-x_{\al^+}$ 
(it necessarily has two jumps symetrically located around $\xj=0$
because of $\Z_2$ antisymmetry):
\be
\f{\al}(\xj)=\sgn(\xj)[g(|\xj|)-\theta(\xj_{\al^+}-|\xj|)]
\label{fdef}
\ee
where $\theta$ is the Heaviside step function. 

Solving for $G$ gives in analogy to (\ref{gdc})
\be
G=dC_{reg}-\frac{\ga}{2}\sum_i h d\xj\we\om_i
-\frac{\ga}{2}\sum_i \f{i} I_i
-\ga\sum_{\al^+} h d\xj\we\theta^{(\al)}_3
-\ga\sum_{\al^+} \f{\al}d\xj\we\de^{(\al)}_4
\label{brgsol}
\ee
where $d\theta^{(\al)}_3=\de^{(\al)}_4$. 
Therefore, we have to sum in (\ref{brgsol}) only over positive $x_\al$ 
(denoted by $\al^{+}$) since the branes at negative values of $\xj$
are automatically taken into account. 
A similar argument as before (integration over a cycle comprising
only one brane) shows that the function $g$ in (\ref{fdef}) is the
same for all branes and also the same as the one defining 
$\f{1}$ and $\f{2}$.

The condition (\ref{BIglob}) that $G$ be globally defined is also
crucial for anomaly cancellation and for 
confining anomaly to the wall or brane. Let us now describe the
anomaly cancellation in the case with five--branes. We will work with
8--forms (and not 6--forms) and will not keep track of possible
local counterterms.

The contribution to the anomaly from the one--loop result is 
\be
\de S^{1-loop}= \sum_{\al} X_{8,\al}\,.
\label{bronel}
\ee
The contribution to the anomaly from the Green--Schwarz mechanism
is 
\be
-\frac1{\ga}\de\int G\we X_7=-\frac1{\ga}\int G\we dX_6^1=
-\sum_{\al} X_{6,\al}^1\,.
\label{brgs}
\ee
The contribution (\ref{brgs}) translated to the 8--form language
cancels the contribution from (\ref{bronel}). Therefore, we expect
that the contribution from the topological term should vanish. Let us
show that (under some condition) this indeed is the case. Since
$\de_{5}$ is invariant under gauge and gravitational transformations,
$\de C$ is the same as before (\ref{dec}):
\bea
\de C&=&dB_{reg}-\frac{\ga}{2} h \sum_i d\xj\we\om_i^1
-\frac{\ga}{2}\sum_i \f{i}d\om_i^1\nn\\
&=&d\left(B_{reg}-\frac{\ga}{2}\sum_i \f{i}\om_i^1\right)+\ga\sum_i
\de^{(i)}\we\om_i^1\,.
\eea
Using (\ref{brbianid}) we get the five--brane contribution to the
anomaly 
\be
\left.
\int\de C\we G\we G
\right|_{fb}
=
2\ga\int\sum_{\al}\left(B_{reg}-\frac{\ga}{2}\sum_i
\f{i}\om_i^1\right)
\we
\left(dC_{reg}-\frac{\ga}{2}\sum_j \f{j} I_j\right)
\we \de_4^{(\al)}.
\ee
Therefore, if we impose conditions for all positions of branes
$(\xj_\al)$: 
\be
B_{reg}(\xj_\al)=\frac{\ga}{2}\sum_i \f{i}(\xj_\al)\om_i^1
\ee
then the contribution to the anomaly from the topological term
vanishes and summing up all contributions the total anomaly does so
as well.

\section{M--theory on $T^5/\Z_2$}

Another interesting orbifold compactification of M--theory is that on 
$T^5/\Z_2$ i.e.\ down to a theory in six dimensions with 16
supercharges \cite{witfb,dasm,flo}. Similarly as in the ten 
dimensional case there is a potential gravitational anomaly. In the
limit of small compactification radius the particle content of this
theory (after imposing $\Z_2$ on the fields) is a chiral supergravity
multiplet (consisting of a graviton, 2 chiral gravitinos and 
5 selfdual twoforms) and five tensor multiplets (each consisting of
an  
antiselfdual twoform, 2 antichiral fermions and 5 scalars).
This theory does not possess vectors so that the potential anomaly
must be purely gravitational. It turns out that the anomaly vanishes
when there are additional 16 tensor multiplets (so called twisted
sector not directly obtained from the compactification).  
$T^5/\Z_2$ has 32 fixed six--planes and the additional matter
multiplets should live on these six--planes. The anomaly from the
small radius theory (untwisted sector only) is equally distributed
among all 32 points so that the anomaly at the $P$-th plane is given
by:  
\be
I_{8,P}^{(1-loop)}=\frac1{32}\left(2I^{(3/2)}-10I^{(1/2)}\right)+
N_P\left(-2I^{(1/2)}-I^{(3-form)}\right)
\ee
where $N_P$ is the number of the twisted sector matter multiplets
living on the $P$-th fixed plane. Using explicit formulae for the
anomalies in six dimensions one gets
\be
I_{8,P}^{(1-loop)}=(N_P-\frac12)X_8\,.
\ee
Sum over $P$ gives the previously mentioned condition $\sum N_P=16$.
Since $N_P$ are integers it is impossible to cancel the anomaly
at any fixed six-plane and to do so one has to modify Bianchi
identities to provide additional source of anomaly inflow via
Green--Schwarz mechanism. If
\be 
dG=\sum_P g_P\de^{(P)}_5
\label{dgp}
\ee
then by the same arguments as before the full anomaly is given by 
\be
I_{8,P}=(N_P-g_P-\frac12)X_8\,.
\ee
To cancel the anomaly the magnetic charges have to be
half--integers satisfying 
\be
g_P=N_P-\frac12\,.
\label{gpnp}
\ee
We will describe below an explicit construction of $G$ satisfying
(\ref{dgp}) in the upstairs approach. We could apply the techniques
worked out in the case of $S^1/\Z_2$ but it is not necessary. 
The $T^5/\Z_2$ orbifold is simpler because the cohomology
condition (\ref{BIglob}) is now replaced by an algebraic relation 
$\sum N_P=16$.

Let us denote $2^5=32$ fixed six--planes by $P(\bar n)$ where 
${\bar n}={(n_7,n_8,n_9,n_{10},n_{11})}$ and $n_k=0,1$. 
The action of $\Z_2$ on the torus $\otimes_k(-\pr_k,\pr_k]$ 
that leaves these 32 points intact is given by $x_i\to -x_i$. Let us
also introduce $5\times 2^4=80$ intervals joining fixed points by
$I({\bar n}^k)$ where ${\bar n}^k={(n_7,\ldots,n_{11})}$ with $n_k$
left out. The interval $I({\bar n}^k)$ is parallel to the $k$-th axis 
and we assume that its orientation is the same as the orientation of
that axis.  

The field $G$ satisfying (\ref{dgp}) and
antisymmetric under $\Z_2$ is given by 
\be
G=dC
+\frac12\sum_k
\sum_{{\bar n}^k}c({{\bar n}^k}){\rm sgn}(x^k)
\delta_4^{({{\bar n}^k})}
\label{Gcnk}
\ee
where
\be
\delta_4^{({{\bar n}^k})}
=
\frac{1}{4!}\epsilon_{kpqrs}\delta\left(x^p-n_p\pi\rho_p\right)
\we\ldots\we
\delta\left(x^s-n_s\pi\rho_s\right)
\ee
and $c({\bar n}^k)$ are constants characterizing $G$ along 
the interval $I({\bar n}^k)$. 
Calculating $dG$ from (\ref{Gcnk}) and comparing with (\ref{dgp}) 
and (\ref{gpnp}) we get 
\be
g_{P(\bar n)}=\sum_k
(-1)^{n_k}c({{\bar n}^k})
=N_P-\frac12\,.
\label{gpsum}
\ee
In the above equation ${\bar n}^k$ has the same components as 
${\bar n}$ but with $n_k$ dropped. 
It is straightforward for any given set of $N_P$ (satisfying $\sum
N_P=16$ i.e. $\sum g_P=0$) to get $c({{\bar n}^k})$
satisfying the above equation. Let us describe explicitly for example 
the ``checkerboard'' configuration \cite{witfb} 
where 16 tensor multiplets are distributed in the most uniform way:
\be
N_P=\frac12\left[1+(-1)^{n_7+n_8+n_9+n_{10}+n_{11}}\right]\,.
\ee
Shrinking any of the radii in the above configuration, any pair of
fixed points along the shrinking direction contributes 1 tensor
multiplet what corresponds to the string limit \cite{witfb,dasm}.
For this choice for $N_P$ it is easy to derive one of possible
sets $c({{\bar n}^k})$:
\be
c({{\bar n}^k})=\frac1{10}(-1)^{n_7+\ldots n_{11}}
\ee
(where the index $n_k$ is left out on the r.h.s.) and check that
(\ref{gpsum}) is satisfied. There are of course other 
$c({{\bar n}^k})$ -- differring by $dC$ in (\ref{Gcnk}) -- that give
the same local cancellation of anomalies.

Other distributions of tensor multiplets among the fixed planes can be
obtained from the ``checkerboard'' configuration by moving those
mutliplets one by one from some fixed point to the neighbouring one 
along interval $I({\bar n}^k)$. 
Such change requires appropriate modification of $c({\bar n}^k)$:
\be
\Delta c({\bar n}^k)
=\pm\frac12
\label{c12}
\ee 
where the sign depends on the ``orientation'' of the exchange.
 
The theory in the presence of five--branes orthogonal to $T^5$ can be 
similarly analyzed. The location of the $\alpha$-th five--brane is
given by a vector with components $x^k_\alpha$ for $k=7,\ldots,11$. 
In the presence of such five--branes the modified Bianchi identities
read 
\be
dG=\sum_Pg_P\de^{(P)}_5+\sum_\alpha \de^{(\alpha)}_5\,.
\label{BI55}
\ee
Because of the $\Z_2$ symmetry the five--branes must come in pairs 
with opposite coordinates: $x^k_\beta=-x^k_\alpha$. In order to
satisfy (\ref{BI55}) the field $G$ must have jumps at the location of
both five--branes. The requirement that $G$ is globally well defined
can be fulfilled only when there are two additional jumps of opposite 
sign located at some fixed six--planes. 
Such negative jumps of $G$ at the fixed planes correspond to removing
two twisted sector matter multiplets from those fixed planes and
transfering them to the so called ``wandering'' five--branes. 
One of the solutions is to put both of those negative jumps for all
pairs of five--branes at the origin $P({\bar n}=(0,0,0,0,0))$. 
If we want to transfer tensor multiplets from some other fixed
six--planes we have first to move those multiplets to the origin using
the procedure described before (\ref{c12}). 
The solution transferring all multiplets from the origin to the branes
is given by the sum of (\ref{Gcnk}) and 
\be
\Delta G = -\frac{1}{2\cdot5!}\sum_\alpha\sum_{pqrst}
\epsilon_{pqrst}{\rm sgn}(x^p)\theta(|x^p_\alpha|-|x^p|)
\delta\left(x^q-x^q_\alpha\right)
\we\ldots\we
\delta\left(x^t-x^t_\alpha\right)\,.
\ee
Now 16 twisted sector tensor multiplets are distributed between fixed
planes and five--branes 
\be
\sum_P N_P + \sum_\alpha 1 =16\,.
\ee
The extremal situation is reached when all tensor multiplets sit on
five--branes and there are no twisted sector multiplets on the fixed
six--planes. 

\section{Conlusions}

We have analyzed the problem of anomaly cancellation in
the upstairs approach. It turned out that for the compactification on
$S^1/\Z_2$ orbifold all final results are sums of contributions from
two walls so the anomaly cancellation can be achieved without any
corrections to the original action. 
$C$ and $G$ fields depend on an arbitrary function 
of $\xj$ but the anomaly cancellation works in the same way for
any choice of this function (however, one should note that for other
purposes some choice may be preferred over the others like for example
in \cite{MNO} where it was shown that the Kaluza--Klein zero modes 
have some specific dependence on the eleventh dimension).
The cancellation of anomalies in the presence of five--branes is
possible but requires
one additional condition on the variation of $C$ under the gauge and
gravitational transformations. In the case of compactification on
$T^5/\Z_2$ 
(with and without five--branes) we have presented in the upstairs
approach an explicit method to derive field configuration for which
the total anomaly vanishes separately at each fixed six--plane. 

The same results can be obtained in the downstairs approach but 
with appropriate boundary conditions replacing the $\Z_2$ symmetry. 
This, however, is less straightforward to implement. 
For example the upstairs approach 
automatically takes into account proper normalization of charges. 
In the case of the $T^5/\Z_2$ compactification the 
number of additional tensor multiplets is always 16 in the upstairs
approach (since two $\Z_2$ symmetric ``wandering
branes'' carry two tensor multiplets) 
while in the
downstairs approach it can be smaller 
(because now there is only a single brane with one tensor multiplet). 
In conclusion, from the point of view of anomaly cancellation, the
upstairs and downstairs approaches are equivalent. In the actual
computation, however, the upstairs approach seems to be more
convenient.

\vspace{1.0cm}
\noindent
{\Large \bf Acknowledgements}
\vskip .5cm
We would like to thank J. Conrad and H.P. Nilles for discussions. 
K.A.M. was partially supported by the Polish KBN grant 2P03B 03715
(1998-2000).
M.O. was partially supported by the Polish grant 
KBN 2 P03B 052 16 (1999-2000).

\end{document}